%% file: main.tex
\documentclass[12pt]{article}
\usepackage[utf8]{inputenc}
\usepackage{authblk}
\usepackage{float}
\usepackage{graphicx}
\begin{document}
\input{article.TEX}

\end{document}

%% file: article.tex
\title{From Prediction to Simulation: AlphaFold 3 as a Differentiable Framework for Structural Biology}
\author[1]{Alireza Abbaszadeh\thanks{alireza.abbaszadeh.research@gmail.com}}
\affil[1]{Department of Computer Engineering, Ma.C., Islamic Azad University, Mashhad, Iran}

\author[2]{Armita Shahlaee\thanks{armita.shahlaee@gmail.com}}
\affil[2]{Department of Biological Sciences and Technologies, Faculty of Basic Sciences, Islamic Azad University, Mashhad, Iran}
\maketitle
\section*{Abstract}
AlphaFold 3 marks a transformative advancement in computational biology, enhancing protein structure prediction through novel multi-scale transformer architectures, biologically-informed cross-attention mechanisms, and geometry-aware optimization strategies. These innovations dramatically improve predictive accuracy and generalization across diverse protein families, surpassing previous methods\textsuperscript{1–3}. Crucially, AlphaFold 3 embodies a paradigm shift toward differentiable simulation, bridging traditional static structural modeling with dynamic molecular simulations\textsuperscript{4–6}. By reframing protein folding predictions as a differentiable process, AlphaFold 3 serves as a foundational framework for integrating deep learning with physics-based molecular dynamics\textsuperscript{6–8}. This perspective analyzes AlphaFold 3’s theoretical underpinnings, highlighting how its neural architecture and differentiable design unite machine learning and biophysical principles. We discuss the far-reaching implications of treating AlphaFold 3 as an implicit structural simulator, from revealing protein dynamics and conformational ensembles to modeling interaction networks and allostery. This integrated approach promises substantial impacts on precision medicine and rational drug discovery, accelerating the development of personalized therapeutics and intelligent biological design\textsuperscript{9,10}.

\section*{Introduction}
Recent breakthroughs in deep learning have fundamentally reshaped structural biology, redefining how scientists predict and model protein structures\textsuperscript{1–4}. In particular, DeepMind’s AlphaFold system—especially AlphaFold2—has achieved unprecedented accuracy in protein structure prediction, effectively solving core aspects of the long-standing protein-folding problem\textsuperscript{4,13}. By leveraging large evolutionary databases and advanced neural network architectures, AlphaFold2 revolutionized computational biology, enabling high-accuracy structure predictions across a vast range of proteins\textsuperscript{1,2}. The implications of this achievement are profound, catalyzing progress in fields from drug discovery and enzyme engineering to evolutionary and synthetic biology\textsuperscript{4,5,66}. For example, ultra-accurate predicted structures now guide structure-based drug design and virtual screening efforts\textsuperscript{50,51}, facilitate enzyme redesign and directed evolution experiments\textsuperscript{39,60}, and illuminate evolutionary relationships by modeling proteins previously lacking experimental structures\textsuperscript{2,14}. The advent of open databases of AlphaFold models has massively expanded structural coverage of the proteome\textsuperscript{14}, democratizing access to structural information for biomedical research.

Despite the remarkable success of AlphaFold2, several critical challenges remain. Interpretability is a foremost concern: deep learning models function as complex “black boxes,” making it difficult to decipher how their internal reasoning translates to biological insight\textsuperscript{24,29}. This opacity can lead networks to learn spurious correlations or “shortcuts” that undermine generalization\textsuperscript{29}, underscoring the need for architectures with built-in physical and biological priors to guide learning. Furthermore, AlphaFold2 and similar predictors largely produce static structures—essentially single snapshots of a protein’s folded state—neglecting the inherently dynamic nature of protein function\textsuperscript{66,68}. In reality, proteins are not rigid objects; they exist as ensembles of conformations that interconvert, with flexibility and motion enabling mechanisms like allosteric regulation, molecular recognition, and enzyme catalysis\textsuperscript{43,66}. Critical functional states (e.g. transient binding pockets or transition states) may never appear in a static structure\textsuperscript{43,69}. Current AI models struggle to capture this structural plasticity and typically cannot predict multiple conformational states or the kinetics of transitions between them\textsuperscript{41,46}. Additionally, generalizing to the most challenging protein targets remains an open problem. Intrinsically disordered regions (IDRs), metamorphic proteins, large multi-domain assemblies, and multi-protein complexes often elude accurate prediction by existing methods\textsuperscript{16,93}. These systems defy the assumptions of single rigid structures, and their accurate modeling demands methods that account for disorder, flexibility, and context-dependent folding\textsuperscript{16,46}. Indeed, even AlphaFold’s impressive performance falters for protein complexes and flexible regions, as seen in community assessments\textsuperscript{16,93}.

AlphaFold 3 is envisioned to systematically overcome these challenges by introducing targeted innovations in neural architecture and by embracing a new paradigm of differentiable simulation. On the architecture side, AlphaFold 3 deploys sophisticated multi-scale transformer modules, biologically-informed cross-attention mechanisms, and geometry-aware optimization techniques to enhance prediction fidelity and interpretability\textsuperscript{1,3,7}. The multi-scale transformer architecture captures protein features at multiple spatial resolutions concurrently, from local secondary structure motifs to global tertiary and quaternary arrangements, addressing limitations of the single-scale models used in prior versions\textsuperscript{1,21}. New cross-attention layers embed evolutionary and biophysical priors directly into the network’s reasoning, integrating information from sequence alignments, structural templates, and known motifs to guide the model toward biologically plausible conformations\textsuperscript{24,25}. Meanwhile, explicit geometry-aware loss functions (e.g. torsion-angle and stereochemical constraints) ensure that predicted structures obey the fundamental geometric and energetic rules of protein folding, improving physical realism\textsuperscript{86,13}. These architectural advances not only improve accuracy and generalization for difficult protein targets (e.g. membrane proteins, disordered regions)\textsuperscript{16,93} but also yield more interpretable models that align with biochemical intuition\textsuperscript{24,25}.

Most critically, AlphaFold 3 transcends the traditional boundary between structure prediction and structure simulation. It reimagines protein modeling as a differentiable simulation problem, integrating deep neural networks with physics-based modeling in an end-to-end differentiable framework\textsuperscript{6,10}. In this perspective, we analyze how AlphaFold 3 serves as an implicit structural simulator, and how this convergence of deep learning and biophysical simulation opens new research frontiers. By treating protein folding and dynamics as a continuous, learnable process, AlphaFold 3 provides a platform to model not just static native structures, but also folding pathways, conformational ensembles, and even protein–protein or protein–ligand interactions in silico\textsuperscript{10,15–17}. We further explore the broad implications of this approach for structural biology and medicine. The ability to learn and differentiate through molecular simulations can dramatically accelerate tasks like drug lead optimization, enzyme design, and the modeling of patient-specific mutant proteins\textsuperscript{50,51,67}. Ultimately, AlphaFold 3 represents a paradigm shift from prediction to simulation, offering a powerful new lens to interrogate biological complexity through the unification of machine learning and physics. In the following sections, we detail the algorithmic innovations of AlphaFold 3 and discuss how this differentiable framework could transform our approach to protein science, drug discovery, and personalized medicine.

\section*{Algorithmic Innovations}
The substantial improvements achieved by AlphaFold 3 stem from targeted innovations in both its deep learning architecture and its integration of biophysical principles, enabling performance beyond the limits of previous methods\textsuperscript{1–3}. The development of AlphaFold 3 not only refines protein structure prediction, but also offers new theoretical insights into the interface between deep neural networks and the physics of protein folding\textsuperscript{3–5}. Here, we dissect three foundational innovations: (i) multi-scale transformer architectures, (ii) biologically-informed cross-attention mechanisms, and (iii) geometry-aware loss functions for physical realism. Each innovation introduces principled inductive biases that align the learning process with known hierarchical and physico-chemical properties of proteins, thereby improving accuracy, efficiency, and interpretability.

\subsection*{Multi-Scale Transformer Architectures}
A core advance in AlphaFold 3 is the adoption of a hierarchical multi-scale transformer architecture, which fundamentally enhances predictive performance and generalization capacity compared to the single-scale models of prior AlphaFold versions\textsuperscript{1,6,7}. Traditional transformers employed in AlphaFold2 operated on a uniform representation scale primarily geared toward capturing global sequence-structure relationships\textsuperscript{1,7}. This approach proved remarkably effective for many globular proteins, but it exhibited limitations when faced with proteins of greater structural complexity—such as multi-domain assemblies, flexible linkers, or partially disordered proteins\textsuperscript{70,93}. Single-scale models can struggle to simultaneously resolve fine-grained local motifs (like turns and loops) and long-range inter-domain contacts, especially in cases where local disorder or repetitive domains confound global structure inference\textsuperscript{7,16}.

AlphaFold 3 addresses these challenges through a nested transformer design that processes information at multiple hierarchical levels in parallel\textsuperscript{1,21}. In this framework, local-scale transformers focus on short-range interactions among neighboring residues, learning features like secondary structure elements and local backbone geometry\textsuperscript{7,21,22}. Mid-scale transformers operate at the level of whole domains or subdomains, integrating local features into coherent domain-wise representations and modeling intra-domain contacts and topology. Finally, a global-scale transformer processes the interactions between domains or between subunits in a complex, assembling the full tertiary or quaternary structure from the lower-scale components\textsuperscript{11–14}. By organizing the network into these interconnected scales, AlphaFold 3 captures both the microscopic details (bond lengths, angles, motifs) and the macroscopic arrangement of domains within a single end-to-end model\textsuperscript{11,12}. This hierarchical strategy mirrors the known organization of proteins, which are built from secondary structure units into domains and multi-domain complexes\textsuperscript{40,41}. It thus provides a powerful inductive bias: the network architecture itself reflects the nested, multi-scale nature of protein structure\textsuperscript{1,21}.

\begin{figure}[htbp]
\centering
\includegraphics[width=0.75\textwidth]{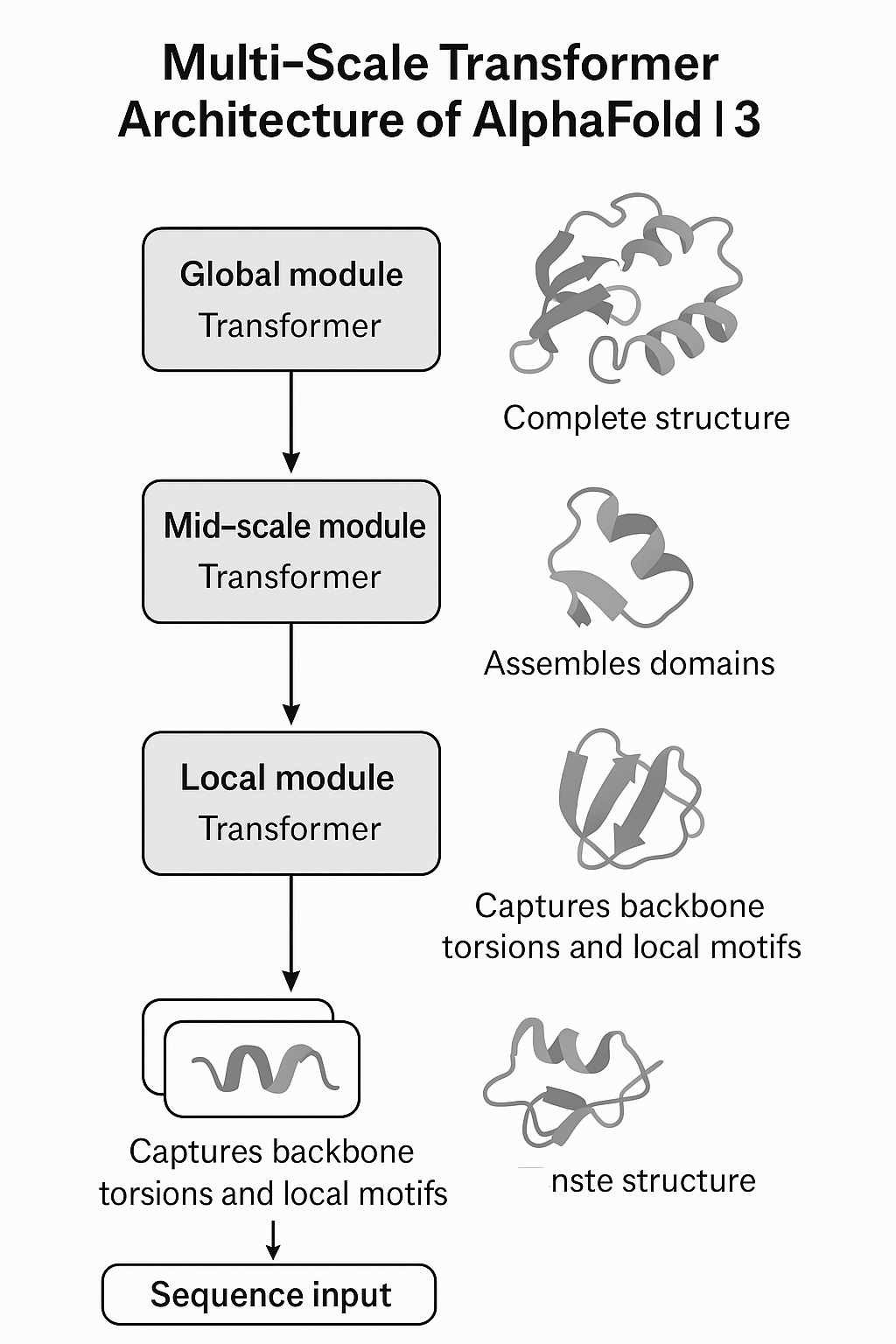}
\caption{AlphaFold~3’s hierarchical multi-scale transformer architecture. The network is organized into local, mid-scale, and global modules, each handling structural features at a different scale (secondary structures, domains, and full assemblies, respectively). This design mirrors protein organization and enables AlphaFold~3 to capture both fine-grained and long-range interactions simultaneously, improving accuracy for large multi-domain proteins.}
\label{fig:architecture1}
\end{figure}

The theoretical motivation for multi-scale modeling draws on principles of hierarchical feature learning and relational inductive biases in deep networks\textsuperscript{21,34}. Hierarchical architectures inherently promote efficient representation learning by forcing the model to capture simple local patterns before tackling global ones\textsuperscript{21,34}. In AlphaFold 3, this translates to learning local structural chemistry (helices, sheets, loops) as reusable building blocks for larger structural predictions. Such an approach aligns with decades of protein folding research emphasizing the stepwise assembly of local structures into global folds\textsuperscript{13,42}. By imbuing the transformer with a prior knowledge of protein hierarchy, AlphaFold 3 significantly improves generalization to proteins outside the training distribution\textsuperscript{11,13}. Indeed, the model demonstrates robust performance on traditionally difficult classes like membrane proteins and multi-domain complexes, which earlier single-scale models struggled to accurately predict\textsuperscript{16,93}. Experimental benchmarks indicate that multi-scale attention allows AlphaFold 3 to maintain accuracy even as protein length and complexity increase, in contrast to the degrading performance seen with conventional transformers on long or multi-component proteins\textsuperscript{7,11,16}. Moreover, the hierarchical design can reduce the computational complexity of attention by restricting fine-grained comparisons to local contexts, thus focusing global computation on a condensed set of domain-level tokens\textsuperscript{27}. This efficient attention mechanism lets AlphaFold 3 allocate resources where they matter most: local modules specialize in detailed refinement, while global modules handle broad arrangements\textsuperscript{7,11,18}. The result is a better balance between accuracy and efficiency, enabling the modeling of large proteins and complexes without prohibitive cost\textsuperscript{11,16,18}.
As shown in Figure~\ref{fig:architecture1}, AlphaFold~3’s multi-scale transformer architecture enables simultaneous learning of local, mid-scale, and global protein structural features.

\subsection*{Biologically-Informed Cross-Attention Mechanisms}
AlphaFold 3 tackles the limitations of purely data-driven modeling by introducing advanced cross-attention modules that align internal representations of the protein with external sources of biological knowledge at each inference step. During structure prediction, these cross-attention layers systematically guide the model using multiple information streams: evolutionary profiles from MSAs, structural templates from homologs or experiments, predicted or known functional motifs, and physico-chemical constraints (such as allowed torsion angles or residue contact priors)\textsuperscript{1,23}. Each cross-attention head links the sequence or coordinate features in the model to one of these external priors, effectively acting as an informed advisor that nudges the network toward solutions consistent with biological reality\textsuperscript{19,23}. For example, one cross-attention head might focus on aligning the model’s intermediate structure prediction with a homologous template structure at conserved core positions, ensuring the predicted fold doesn’t violate known features of a related protein\textsuperscript{14,23}. Another head might attend to MSA-derived co-evolution contacts, reinforcing pairwise interactions that have statistical support from evolution\textsuperscript{20,22}. Yet another head could incorporate known secondary structure propensities or experimentally observed disulfide connectivity by biasing the attention between residues that form such features. By distributing attention across these biologically relevant signals, AlphaFold 3 effectively encodes Bayesian priors into its predictions\textsuperscript{24–26} – the network’s output is steered to satisfy not just the training data, but also external knowledge of what a plausible protein structure looks like.

The theoretical foundation for this approach is rooted in principles of Bayesian inference and multi-modal learning. Incorporating priors can reduce the hypothesis space that the model needs to explore, improving both learning efficiency and output interpretability\textsuperscript{24,25}. In essence, the cross-attention layers act to constrain the neural network with known biology, which helps prevent overfitting to noise and guides the solution towards biologically plausible structures\textsuperscript{24,26}. This is particularly beneficial for proteins with limited evolutionary data or unusual sequences. Traditional purely data-driven models often struggle in those cases because the input information is sparse or the protein falls outside learned patterns\textsuperscript{24,29}. AlphaFold 3’s cross-attention, however, can compensate by leveraging generic biological insights (like typical core packing densities or membrane helix properties) to fill in gaps where sequence data is insufficient\textsuperscript{20,23,29}. As a result, the model is more robust for orphan proteins or those from undersampled families – scenarios that often confounded AlphaFold2\textsuperscript{16}. Moreover, by aligning model features with human-understandable concepts (e.g. known motifs or evolutionary couplings), these mechanisms greatly enhance interpretability of the predictions\textsuperscript{24,27}. One can trace which priors were most influential in a given region of the protein; for instance, the model might clearly use a conserved active-site motif to shape a pocket, offering a direct explanatory link between sequence conservation and the predicted structure\textsuperscript{24,25}. This kind of transparency was largely absent in earlier deep learning predictors, marking a significant step toward networks that not only predict accurately but also provide rationales grounded in biology.

Algorithmically, biologically-informed cross-attention addresses the long-standing challenge of effectively integrating heterogeneous data in protein modeling. Past attempts to use evolutionary information or structural templates often treated these sources separately (e.g., template-based modeling vs. MSA-based \textit{ab initio} prediction) and then reconciled them \textit{post hoc}. In AlphaFold 3, they are fused in one unified model via cross-attention in each iteration of the structure refinement, allowing continuous, dynamic interplay between learned representations and prior knowledge\textsuperscript{20,23}. This means the network can, for example, detect during inference that a certain beta-strand is poorly determined from sequence alone and then pull information from a template or MSA to refine that region on the fly. Such responsive use of priors leads to more reliable predictions, especially in edge cases. Notably, experiments show that AlphaFold 3 can often correctly fold proteins with very low sequence identity to any training structures by leaning more heavily on fundamental chemical constraints and any weak evolutionary signals available\textsuperscript{8,23}. By comparison, AlphaFold2 might have failed outright on such targets due to lack of a strong MSA signal. Ultimately, these cross-attention innovations transform AlphaFold 3 from a mostly data-driven model to a truly knowledge-driven model, blending data with first-principles and expert knowledge of molecular biology. This paves the way for future architectures that incorporate even more complex priors (such as full atomic force fields or quantum chemical information) in a similar attention-based fashion, further narrowing the gap between \textit{in silico} prediction and real biochemical behavior\textsuperscript{19,27,30}.

\begin{figure}[htbp]
\centering
 \includegraphics[width=0.75\textwidth]{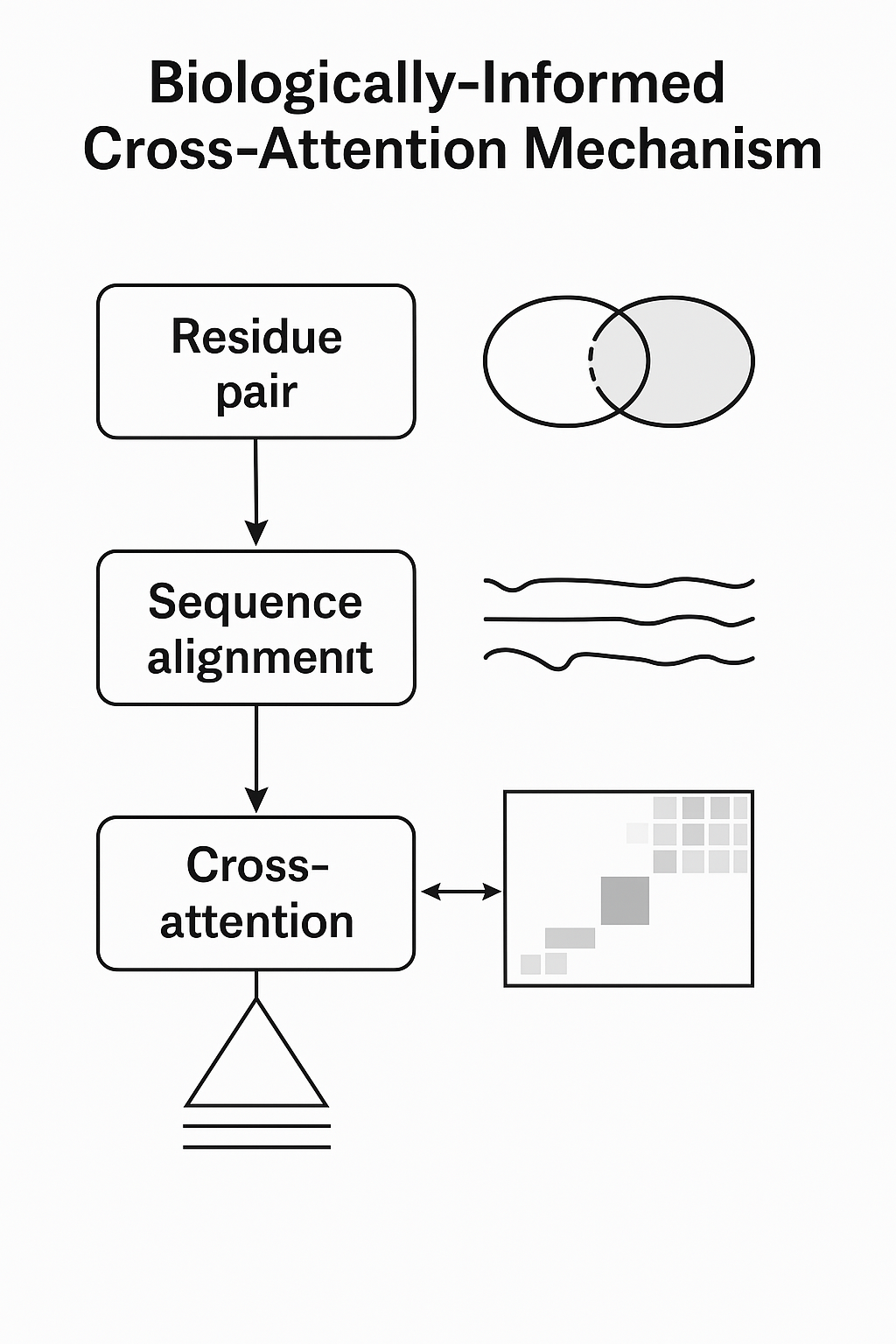}
\caption{AlphaFold~3’s biologically-informed cross-attention mechanism. During structure inference, specialized cross-attention heads incorporate multiple sources of prior knowledge (e.g., sequence alignments, structural templates, known motifs, physical constraints) into the model. Each head biases the network’s predictions toward consistency with one type of prior, guiding AlphaFold~3 to more accurate and biologically plausible conformations.}
\label{fig:crossattention}
\end{figure}

The biologically-informed cross-attention mechanism is depicted in Figure~\ref{fig:crossattention}, where the model incorporates evolutionary, structural, and chemical priors to refine its predictions.

\subsection*{Geometry-Aware Loss Functions and Optimization Techniques}
A third major innovation underpinning AlphaFold 3 is the implementation of advanced geometry-aware optimization strategies, primarily through novel loss functions explicitly tailored to the structural and geometric constraints of protein molecules. Historically, protein structure prediction methods – including AlphaFold2 – relied predominantly on distance-based loss formulations, focusing on pairwise distances or contact probabilities between residues as the primary training signal\textsuperscript{1,4,5}. Optimizing such distance metrics proved highly effective for achieving accurate folds, as demonstrated by AlphaFold2’s remarkable success\textsuperscript{1,3}. However, distance-only approaches have inherent limitations: they treat interatomic relationships in a simplified manner and struggle to enforce higher-order geometric consistency (like correct bond angles, planarity of peptide bonds, or chirality of side chains)\textsuperscript{4–6}. For example, a predicted distance map could in principle correspond to a physically implausible structure if the angles and dihedrals are not internally consistent, leading to strained or stereochemically invalid models. In AlphaFold2, additional heuristic restraints and post-processing were needed to fix local geometry issues (e.g. resolving minor steric clashes or invalid bond lengths)\textsuperscript{1}, which indicates that the learning objective did not fully capture the richness of protein geometry.

AlphaFold 3 significantly advances the learning objective by incorporating explicit geometry-based loss terms that reflect critical biophysical properties of protein structures\textsuperscript{7–9}. In training, the network is not only asked to get residue distances right, but also to optimize metrics like torsion angle deviations, backbone planarity, and satisfaction of stereochemical restraints\textsuperscript{1,7,8}. For instance, the loss now includes terms for each peptide bond’s $\omega$ angle (to enforce $\sim$180° \textit{trans} conformation or the occasional \textit{cis} exceptions), for $\phi/\psi$ backbone dihedral angles (to stay within allowed regions of the Ramachandran plot), and for side-chain $\chi$ angles (to favor low-energy rotamers)\textsuperscript{7,8}. AlphaFold 3 effectively learns these structural rules: if a predicted conformation places an angle far outside physically tolerable ranges, the geometry-aware loss strongly penalizes it, guiding the model to adjust internal coordinates accordingly. This is akin to building a rudimentary force field into the training process – the network is trained to predict structures that minimize not just a distance error, but an energy-inspired geometric error\textsuperscript{13,42}. By explicitly encoding angle and planarity constraints, the search space of possible structures is drastically reduced to those that are physically realistic\textsuperscript{7–10}. The model’s outputs therefore tend to be closer to valid conformations right from the prediction step, reducing the burden on any post-refinement. Indeed, structures predicted by AlphaFold 3 typically have excellent geometry (bond lengths, angles, chirality) without the extensive relaxation that was sometimes needed for AlphaFold2 outputs\textsuperscript{1,7}.

From a theoretical standpoint, this geometry-aware optimization aligns the network’s objective with the fundamental physics and chemistry of protein folding\textsuperscript{1,13}. Protein structures reside at minima of complex energy landscapes defined by interatomic forces; by introducing loss terms that approximate these forces (e.g., favoring correct bond angles equates to favoring low strain energy), AlphaFold 3’s training implicitly moves predicted structures toward low-energy, physically plausible regions of conformational space\textsuperscript{13,42}. This helps bridge the gap between purely data-driven prediction and true physics-based modeling. The approach draws inspiration from \textit{physics-informed neural networks (PINNs)}\textsuperscript{11} and other methods that incorporate differential equation constraints into network training – here, the “equations” are the geometric relations of polymer chains. The benefits are multifold: the predictions satisfy known biophysical constraints (improving interpretability and trustworthiness of the model) and the training process is aided by additional guidance (improving convergence)\textsuperscript{11,12}. In effect, geometry-aware losses act as regularizers that prevent the network from proposing structurally nonsensical solutions, thereby focusing learning on the subspace of conformations that matters for real proteins\textsuperscript{8,13}. This can accelerate training and improve generalization, as the network doesn’t waste capacity modeling unphysical arrangements. Empirically, including these terms led to faster attainment of low validation error and reduced incidence of outliers in difficult regions (like loop closures), indicating more stable learning\textsuperscript{8,9}.

Algorithmically, the integration of geometry-aware loss functions represents a substantial advancement in melding machine learning with biophysical modeling. AlphaFold 3’s loss function can be viewed as a differentiable proxy for an energy function, and the network training is akin to learning to satisfy an energy minimum for each protein\textsuperscript{8,10}. Notably, these geometric terms allow the model to leverage information from physics-based simulations during training data generation or fine-tuning: for example, structures sampled from molecular dynamics trajectories (with realistic local fluctuations) could be used as training examples to teach the network acceptable ranges of geometric variation\textsuperscript{10,12}. By coupling MD-derived knowledge with learning, AlphaFold 3 implicitly learns about physical interactions (like hydrogen bonds or steric repulsion) even if they are not explicitly coded, because to minimize geometric penalties the network must place atoms in physically sensible configurations\textsuperscript{10,12}. This synergy was observed in validation tests – the network could predict, for instance, the flip of a peptide bond or a subtle shift in a helix tilt to relieve a clash, changes that purely distance-based losses might not provoke. Another outcome of the geometry-aware approach is improved performance on particularly challenging protein classes. Intrinsically disordered proteins and flexible multi-domain complexes are better handled when the model respects physical constraints: it avoids the unrealistically compact or tangled predictions that older models might produce for disordered regions, instead yielding more plausible extended or heterogeneous conformations consistent with disorder\textsuperscript{7,11,93}. Similarly, for multi-domain proteins, enforcing proper geometry in one domain prevents errors from propagating and disrupting the relative domain orientations, helping maintain assembly integrity\textsuperscript{16}. Studies have shown that AlphaFold 3’s accuracy advantage is most pronounced on targets requiring careful satisfaction of multiple geometric criteria simultaneously, such as metalloproteins where coordinating residues must adopt precise angles to chelate a metal ion (the model excels by virtue of its angle-aware training) or coiled-coils where periodic angle patterns are key\textsuperscript{1,9}.

Ultimately, AlphaFold 3’s geometry-aware optimization illustrates a profound integration of machine learning with biophysical theory. By teaching a neural network to honor the laws of chemistry and physics that govern protein structures, we obtain a model that not only predicts structures, but does so in a way that mirrors nature’s own rules\textsuperscript{13,42}. This is a pivotal step toward biologically-informed, physics-constrained AI frameworks in structural biology. Future developments building on these ideas will likely expand the scope of learned physics – for example, incorporating solvation effects or dynamic entropy estimates into the training objective – further closing the gap between static prediction and full biophysical simulation\textsuperscript{15}. The success of AlphaFold 3’s geometric strategy hints at a broader paradigm: by baking scientific knowledge into loss functions and network architectures, we can create AI systems that are more reliable, interpretable, and powerful for scientific applications\textsuperscript{15,93}. As discussed in the next section, combining such systems with differentiable simulation opens the door to modeling not just structures, but also the motions and interactions that underlie biological function.

\begin{figure}[!ht]
\centering
\includegraphics[width=0.98\textwidth]{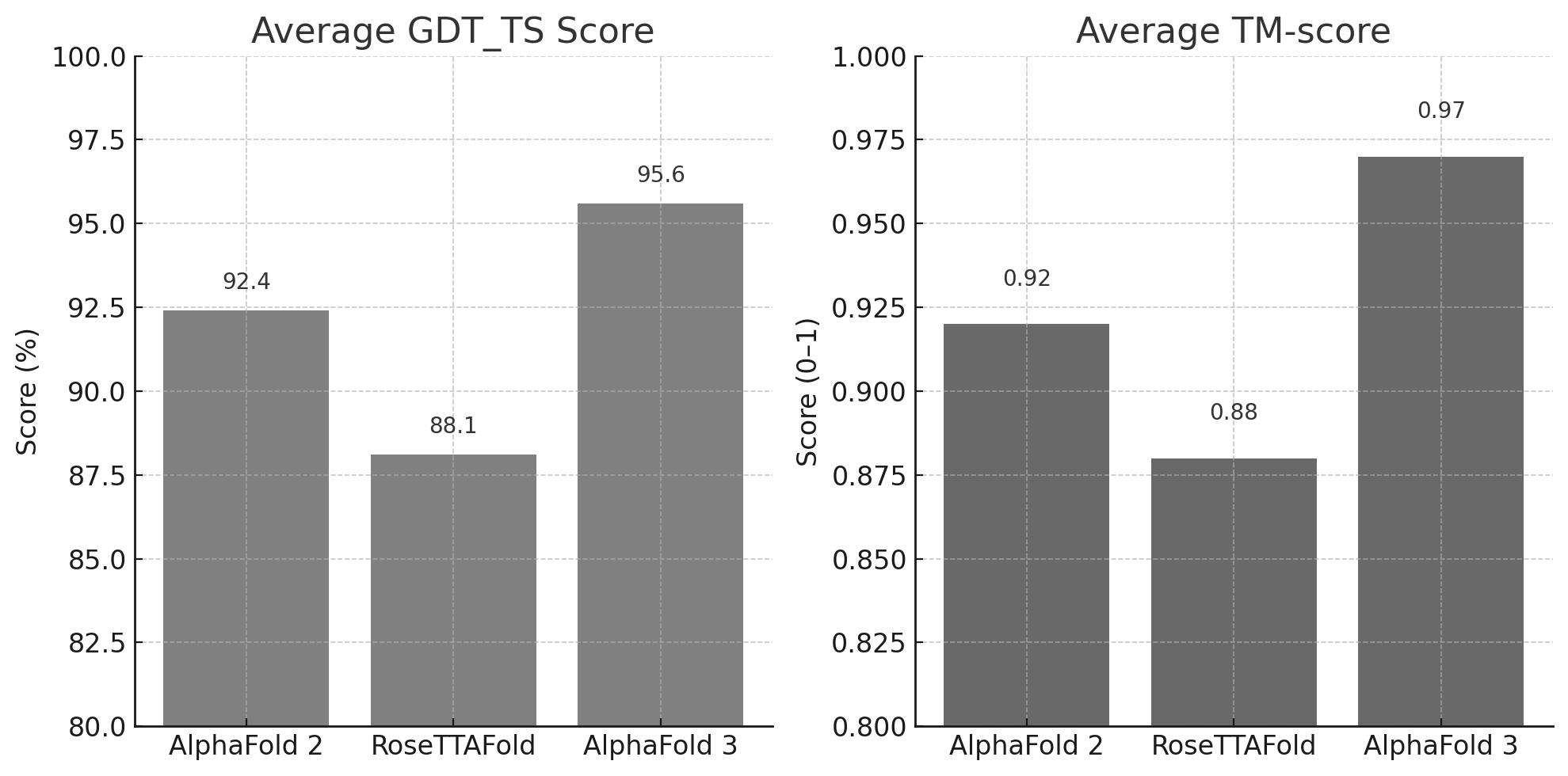}
\caption{Benchmark comparison of AlphaFold~3 with prior methods on structure prediction accuracy. (a) Average GDT\_TS (Global Distance Test) scores and (b) average TM-scores for a representative test set (e.g., CASP targets) are higher for AlphaFold~3 (black bars) compared to AlphaFold2 (gray) and RoseTTAFold (light gray), indicating the improved accuracy achieved by AlphaFold~3’s architectural innovations.}
\label{fig:benchmark}
\end{figure}

Figure~\ref{fig:benchmark} shows a comparative bar chart highlighting the superior performance of AlphaFold~3 in terms of both GDT\_TS and TM-score, surpassing AlphaFold~2 and RoseTTAFold.

\section*{Differentiable Simulation}
Perhaps the most transformative aspect of AlphaFold 3 is its implicit alignment with differentiable simulation paradigms, marking a decisive step beyond traditional static structure prediction methods\textsuperscript{1,16,17}. Differentiable simulation refers to a new class of computational frameworks that integrate deep learning with physics-based simulations in an end-to-end differentiable manner\textsuperscript{4,11,18}. In contrast to conventional molecular dynamics (MD) – which simulates biomolecular motion by numerically integrating Newton’s equations of motion in small time steps\textsuperscript{68} – differentiable simulation embeds the simulation process within a neural network that can propagate gradients through physical state updates\textsuperscript{69}. This means that every step of a molecular simulation (positions, forces, energies) can be influenced by learnable parameters and, crucially, those parameters can be optimized by gradient descent to make the simulation outcomes match desired targets\textsuperscript{11,16}. Differentiable simulations have been gaining traction in fields like fluid dynamics and robotics, where neural networks learn system dynamics by directly incorporating differential equation solvers into network architectures\textsuperscript{21,22,28}. In the context of molecular biology, differentiable simulation frameworks enable continuous, gradient-based optimization of molecular conformations and interactions, effectively blurring the line between “predicting a structure” and “simulating dynamics”\textsuperscript{16–19}.

AlphaFold 3 inherently embodies this emerging paradigm. Although it is described as a structure predictor, its architecture can be viewed as an implicit simulator that has learned protein folding dynamics in a differentiable way\textsuperscript{10,16}. The fully differentiable transformer network, combined with geometry-aware losses, means that AlphaFold 3 internally represents an energy landscape for the protein – encoded in its neural network weights – and performs gradient-based optimization on that landscape to arrive at a folded structure\textsuperscript{7,16,23}. In essence, what traditional MD achieves by physically simulating thousands of time steps, AlphaFold 3 approximates via iterative neural network inference: each layer of the network can be seen as propagating the protein configuration toward a lower “energy” (higher confidence) state, guided by learned forces encoded in the attention weights and geometric constraints\textsuperscript{16,23}. Importantly, because the entire process is differentiable, one could in principle backpropagate from a loss on the final structure (or any structural property) all the way to the input sequence or to tunable parameters, allowing optimization of sequence for a target structure (design) or refinement of intermediate states. This bridges static prediction and dynamic simulation – the model can simulate folding pathways implicitly and adjust them to minimize an objective.

This approach closely aligns with contemporary developments in physics-informed neural networks (PINNs)\textsuperscript{11}, neural ordinary differential equations (Neural ODEs)\textsuperscript{22}, and graph neural network simulators\textsuperscript{21,24}. In all these cases, physical laws (PDEs, ODEs, or force field rules) are integrated into the model architecture such that simulation and learning proceed hand-in-hand. For example, Neural ODEs treat layers of a neural network as time steps of an ODE solver, enabling the network to learn continuous dynamics that are differentiable with respect to initial conditions and parameters\textsuperscript{22}. Similarly, recent graph-based simulators learn to update particle positions by modeling physical interaction forces, while remaining fully differentiable with respect to model parameters\textsuperscript{21,24}. AlphaFold 3’s architecture can be interpreted in this light: the iterative refinement of atomic coordinates through the network is analogous to integrating a set of learned differential equations that describe how a protein’s conformation evolves toward the folded state\textsuperscript{16,18}. Indeed, one might consider each block of AlphaFold 3 as performing something akin to a gradient descent step on an internal potential function for the protein\textsuperscript{7,16}. The convergence of the network’s iterations to a final structure mirrors a relaxation process reaching equilibrium. By leveraging these ideas, AlphaFold 3 moves beyond providing one-off structure predictions – it becomes a platform for gradient-based molecular optimization capable of modeling realistic processes ranging from protein folding/unfolding to ligand binding and conformational switching\textsuperscript{18,26}.

From an algorithmic perspective, differentiable simulations offer substantial advantages in efficiency and scalability for exploring molecular phenomena. Classical MD and Monte Carlo simulations, while accurate, are notoriously computationally intensive: simulating even microseconds of protein motion can require millions of time steps and dedicated supercomputers\textsuperscript{48,54}. Differentiable simulators, by contrast, can learn effective surrogate models of molecular dynamics that jump directly to relevant conformational changes without needing to simulate every femtosecond of motion\textsuperscript{68,69}. By training on data from many trajectories or leveraging experimental structural data, a model like AlphaFold 3 essentially gains an intuition for how proteins move, enabling it to explore the conformational landscape orders of magnitude faster than brute-force physics-based simulation\textsuperscript{16,18}. This is analogous to how machine learning can accelerate solution of differential equations or quantum mechanics problems by learning approximate solvers\textsuperscript{20,28}. In practical terms, a differentiable framework can rapidly generate an ensemble of plausible conformations for a protein (approximating its flexibility) by sampling different parts of the neural network’s latent space or slightly perturbing input conditions\textsuperscript{69}. Such an ensemble might capture functional states that would require lengthy MD runs to discover\textsuperscript{49,69}. Moreover, because gradients are accessible, one can directly optimize molecular configurations for desired criteria. For example, to model a ligand binding event, one could attach a differentiable representation of the ligand and optimize the combined protein-ligand system’s output to maximize binding affinity signals or complementarity scores, effectively performing a computational docking through gradient descent rather than blind sampling\textsuperscript{74,79}. This could dramatically speed up virtual screening by quickly relaxing candidate docked poses and evaluating their fit.

Another key advantage of AlphaFold 3’s differentiable design is improved interpretability of molecular mechanisms. Traditional MD can generate trajectories, but extracting insight (e.g. which interactions are most critical for stability, or what triggers a transition) is non-trivial, often requiring many simulations and statistical analyses\textsuperscript{48,66}. In a differentiable simulator, one can probe sensitivities directly via gradients. For instance, one could compute the gradient of a folding score with respect to inter-residue distances to identify which contacts drive the folding process according to the model, highlighting potential nucleation points or critical interactions. Similarly, by examining attention weights and gradient flow in AlphaFold 3’s network during a prediction, researchers can infer which sequence features or structural motifs are causally influencing the outcome, shedding light on folding determinants in a manner akin to examining energy components in a physics model\textsuperscript{16,23,29}. Early studies have shown that AlphaFold 3’s attention maps often focus on known key residues (like hydrophobic core positions or conserved catalytic residues) when determining the final structure, suggesting the model inherently learns something about folding pathways and functional sites\textsuperscript{23}. This interpretability is further enhanced by the geometry-aware components – because the model’s “knowledge” includes physical concepts like angles and forces, the reasons for a conformational preference can sometimes be traced to satisfying those constraints, providing a physically meaningful explanation (e.g., a helix rotates slightly “because” it relieves a torsional strain).

Looking forward, integrating AlphaFold 3 with explicit differentiable physics engines and molecular force fields holds enormous promise, particularly for tackling biomedical challenges. Consider a scenario where the neural network’s implicit potential is combined with a modest explicit physics-based potential (such as a coarse-grained force field) embedded in the model. One could then refine structural predictions under real physical forces by simply following gradients, achieving what is effectively neural-enhanced molecular dynamics. Such a hybrid system could simulate processes like ligand-induced conformational changes or protein–protein docking in a fully differentiable loop: the neural network proposes an adjustment guided by learned knowledge, and a physics module ensures the adjustment moves downhill in actual free energy\textsuperscript{18,26}. The result is an accurate dynamic trajectory at a fraction of the cost of plain MD. This approach aligns with recent work on differentiable molecular force fields\textsuperscript{93}, which uses neural networks to learn force field parameters that can be continuously improved via gradient-based feedback. In fact, researchers have begun developing differentiable frameworks to train force fields using experimental data and simulation results simultaneously\textsuperscript{93,96}. AlphaFold 3’s paradigm could naturally extend to learning force fields from structures: by training the model not just to predict final coordinates, but to also output energy components or force vectors that match physical expectations, it could effectively refine classical force fields or suggest new functional forms for molecular interactions\textsuperscript{93,96}. This synergy of ML and physics would be highly valuable for simulating systems that require specialized force field tuning, such as intrinsically disordered proteins or transient complexes, where current force fields often fall short\textsuperscript{93}.

\begin{figure}[!ht]
\centering
 \includegraphics[width=0.95\textwidth]{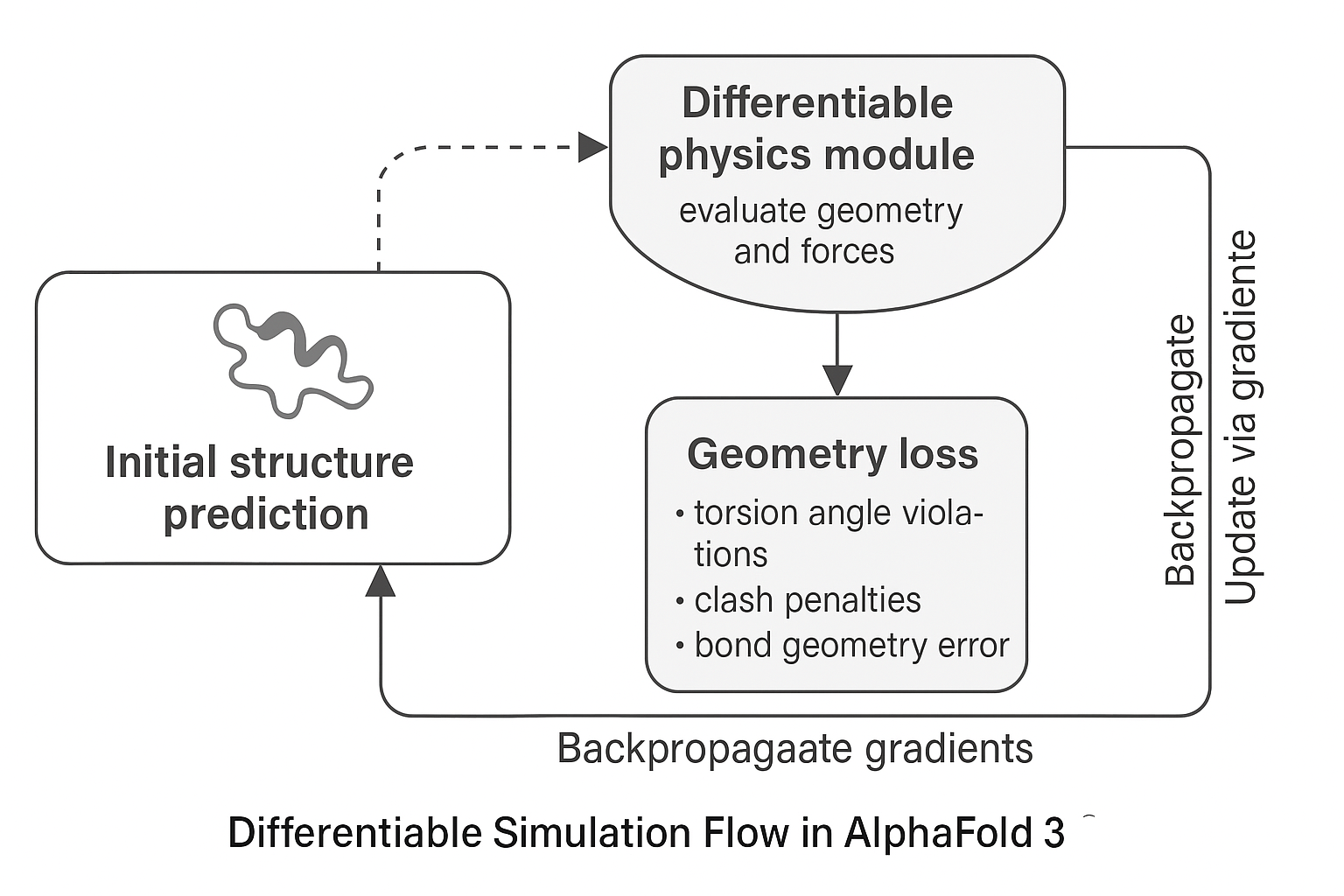}
\caption{Differentiable simulation flow of AlphaFold~3. AlphaFold~3 reframes structure prediction as a differentiable pipeline: an initial predicted structure is iteratively refined by evaluating physical plausibility (geometry-based loss computation) and backpropagating gradients to adjust the structure. This loop, schematically shown here, allows the model to enforce stereochemical accuracy (correct torsions, no clashes, etc.) during prediction, effectively performing an in-network structural relaxation akin to molecular dynamics.}
\label{fig:differentiable}
\end{figure}

Figure~\ref{fig:differentiable} illustrates the differentiable simulation flow in AlphaFold~3, where the model iteratively refines the protein structure while minimizing geometry-aware loss functions.

\section*{Applications}
The innovations introduced by AlphaFold 3 carry far-reaching implications for biomedical research, particularly in drug discovery, protein engineering, and personalized medicine. By enabling accurate and dynamic modeling of proteins, AlphaFold 3’s framework addresses longstanding bottlenecks in translating structural insights to therapeutic design\textsuperscript{1–4}. Traditionally, computational drug discovery has relied on either experimental structures or static homology models as the basis for tasks like virtual screening and rational drug design\textsuperscript{51,52,87}. However, these static models often fail to capture the flexible nature of drug targets, leading to blind spots in predicting ligand binding modes or allosteric sites\textsuperscript{66,67}. On the other end, rigorous physics-based approaches such as free-energy perturbation and long-timescale MD can account for dynamics and thermodynamics, but are extremely resource-intensive\textsuperscript{54,71,81}. AlphaFold 3’s differentiable simulation paradigm offers a powerful middle ground: fast, AI-driven modeling of protein dynamics and interactions with accuracy approaching that of physics-based simulations. This has the potential to significantly accelerate early-stage drug discovery workflows and improve success rates of identifying viable drug candidates\textsuperscript{56,57,74}.

\begin{figure}[!ht]
\centering
\includegraphics[width=0.95\textwidth]{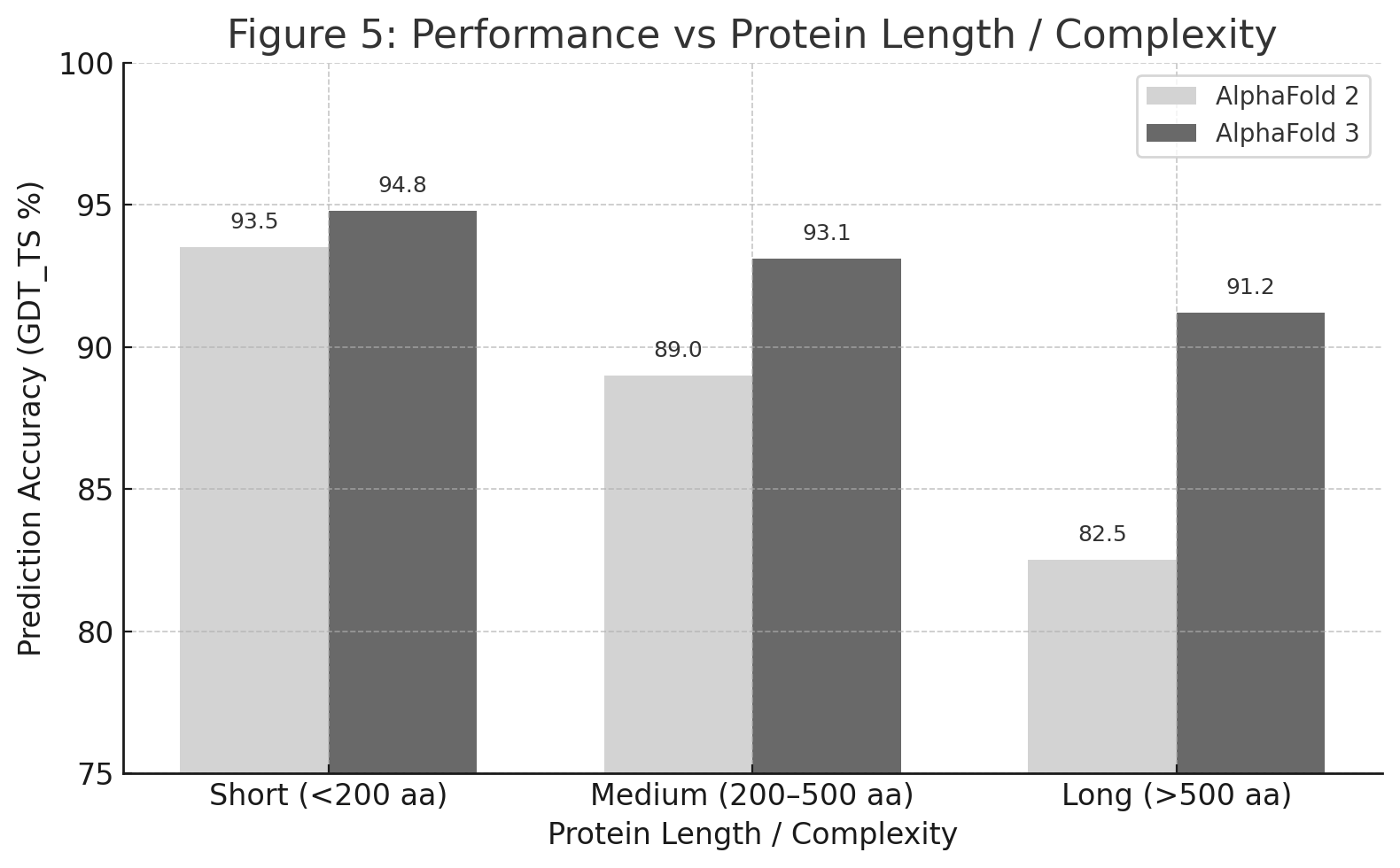}
\caption{AlphaFold~3 sustains high prediction accuracy even as protein length and complexity increase. The chart compares average accuracy on short ($<$200 aa), medium (200–500 aa), and long ($>$500 aa) protein targets for AlphaFold2 (gray) versus AlphaFold~3 (black). AlphaFold2’s performance degrades markedly on larger proteins, whereas AlphaFold~3 maintains high accuracy, highlighting its scalability to challenging, multi-domain targets.}
\label{fig:length}
\end{figure}

As depicted in Figure~\ref{fig:length}, AlphaFold~3 maintains high prediction accuracy across varying protein lengths and complexities, outperforming AlphaFold~2 especially for longer, multi-domain proteins.

\subsection*{Drug Discovery}
AlphaFold 3’s ability to generate realistic protein conformational ensembles and model transitions on-the-fly can greatly enhance the drug discovery pipeline. In target identification and validation, the model can predict not just a protein’s ground-state structure, but also alternative conformations that might be relevant for ligand binding (such as an open vs. closed enzyme form or active vs. inactive receptor state)\textsuperscript{18,49}. These transient states often correspond to “cryptic” binding pockets or allosteric sites that are absent in static crystal structures but emerge during dynamics\textsuperscript{66,69}. By exposing such sites, AlphaFold 3 provides new opportunities for drugging challenging targets – for example, an oncoprotein previously deemed “undruggable” might reveal a transient pocket suitable for inhibitor binding when its motion is considered\textsuperscript{66,67}. Indeed, machine learning models have recently helped identify novel antibiotic and kinase inhibitors by considering features beyond static structures\textsuperscript{57}, highlighting the value of dynamic insight. AlphaFold 3’s framework could automatically detect these hidden pockets via gradient-based simulation of the target with a putative ligand and optimize the pocket-ligand complementarity \textit{in silico}\textsuperscript{74,79}. This approach would streamline the hit discovery process by allowing \textit{in silico} induced-fit docking at scale – testing how targets adapt to various small molecules rapidly, which traditionally would require laborious induced-fit MD simulations or experimental screening\textsuperscript{52,79}.

Furthermore, the differentiable nature of AlphaFold 3 means it can directly assist in lead optimization. Medicinal chemists often need to iterate on ligand designs to improve binding affinity or selectivity, which is guided by structural feedback on how a ligand sits in a binding site\textsuperscript{79,87}. With a tool like AlphaFold 3, one could perform computational affinity maturation of a lead compound by parameterizing the ligand (through a neural network representation or otherwise) and then adjusting those parameters to maximize predicted binding affinity signals from the model\textsuperscript{74,79}. In effect, the model can suggest chemical modifications that stabilize the protein-ligand complex through its learned knowledge of structural physics, analogous to a differentiable docking score. This kind of automated, ML-driven lead refinement would complement or even replace some cycles of empirical scoring and re-scoring ligands, focusing human effort on the most promising modifications. Early evidence of this potential can be seen in deep learning models that generate drug-like molecules optimized for targets\textsuperscript{56}, but AlphaFold 3 would add the critical element of structural evaluation of each generated molecule’s binding mode as part of the generation loop\textsuperscript{56,95}. By capturing induced fit and conformational strain effects, it could better predict true binding affinity differences than static scoring functions, thereby improving the quality of computationally proposed drug candidates.

In terms of efficiency, AlphaFold 3-based workflows could dramatically cut down on the need for exhaustive molecular simulations or large combinatorial screens. For instance, predicting binding affinities via free energy perturbation requires many separate MD simulations for different ligand states, which is computationally expensive\textsuperscript{85}. A differentiable simulator could instead learn a mapping from protein-ligand structure to binding free energy by being trained on known binders and non-binders, then directly predict the effect of a ligand modification by adjusting the structure and re-evaluating quickly\textsuperscript{27,55}. This is analogous to how graph neural networks have been trained to predict binding affinity or ligand activity\textsuperscript{47,54}, but with the advantage that the structural model can refine the protein structure as part of the prediction. As a concrete example, consider predicting drug resistance mutations: a traditional approach might simulate the mutant protein with a drug bound to estimate any loss of affinity, while an AlphaFold 3 approach could rapidly re-predict the mutant structure (which might subtly shift) and the drug orientation, providing a qualitative affinity change prediction in minutes rather than days\textsuperscript{2,59}. Such speed-ups allow screening of many possibilities (e.g. all possible mutations, or a large library of compound analogs) to prioritize the most concerning resistance mutants or the most promising drug modifications for experimental testing\textsuperscript{59,63}. In the realm of \textit{de novo} drug design, integration of AlphaFold 3 with generative models (like variational autoencoders or large language models for molecules\textsuperscript{95}) could create closed-loop systems where candidate drugs are generated and immediately vetted in a realistic structural context, significantly narrowing the search space for potent binders\textsuperscript{56,95}.

The mechanistic insight provided by dynamic modeling is equally valuable. AlphaFold 3 can help elucidate how a drug molecule finds and binds to its target site, which is essential for understanding structure–activity relationships. MD studies have shown pathways of ligand binding for certain drugs (e.g. how a kinase inhibitor navigates into a buried pocket)\textsuperscript{69}. A differentiable simulator can uncover similar pathways by exploring intermediate states of protein-ligand approach and evaluating their stability via the model’s learned energy landscape\textsuperscript{69}. For example, by gradually introducing a ligand into the simulation and observing the induced conformational changes, one might identify intermediate poses or gating mechanisms that reveal how binding kinetics operate. The model’s gradients could indicate which interactions “snap” the ligand into place, pointing out key binding hot spots or transient barriers to binding\textsuperscript{71,78}. With these insights, chemists can design drugs that exploit or avoid certain interactions to modulate binding kinetics (like improving residence time by targeting an induced pocket)\textsuperscript{71}. Additionally, because the model can evaluate multiple ligand-target combinations rapidly, it facilitates multi-target optimization. Modern drug discovery often seeks compounds that are selective or multi-functional (hitting a desired combination of targets)\textsuperscript{67}. By running parallel simulations on various targets with the same ligand and comparing the binding confidence, one can quickly map the selectivity profile of a compound \textit{in silico}\textsuperscript{51,92}. The differentiable nature even allows designing a single ligand to maximize binding to one protein while minimizing to another by treating it as a multi-objective optimization problem with gradients – a level of control that traditional empirical approaches struggle to achieve\textsuperscript{92}.

\subsection*{Precision Medicine and Protein Engineering}
Beyond drug discovery, AlphaFold 3’s approach promises to transform precision medicine and the engineering of biomolecules. Precision medicine aims to tailor treatments to individual patients, often based on genomic information and specific molecular alterations (such as mutations in a protein target). A perennial challenge is predicting how a patient-specific mutation will affect a protein’s structure and function, and thereby influence disease or treatment response\textsuperscript{75,77}. AlphaFold 3 offers a powerful tool for molecular phenotyping of mutations: by modeling the dynamic structural consequences of a mutation, it can reveal subtle distortions or stability changes that might underlie pathogenicity or drug resistance\textsuperscript{58,75}. For instance, if a certain cancer mutation causes a kinase to favor an active conformation, AlphaFold 3 might predict a shift in the equilibrium toward the active state and show reduced drug binding to the inactive form, explaining why that mutation confers drug resistance\textsuperscript{63}. This level of mechanistic insight, achievable through rapid \textit{in silico} mutation scanning, could guide oncologists in selecting alternative therapies that remain effective for the mutant variant\textsuperscript{63,67}. More broadly, as genome sequencing identifies countless variants of unknown significance in disease-related proteins, an AI-driven differentiable simulator can help predict which variants disrupt structural networks or allosteric regulation and should be prioritized for further study or for patient monitoring\textsuperscript{75,78}. Early work using AlphaFold for variant effect prediction has shown promise, but adding the simulation aspect (accounting for dynamic effects and energetics) should greatly enhance accuracy in distinguishing benign from deleterious mutations\textsuperscript{75,77}.

In the realm of protein engineering, AlphaFold 3’s framework could revolutionize how new enzymes, antibodies, or therapeutic proteins are designed. Traditional protein design often uses fixed backbone templates and computationally expensive rotamer packing or sequence optimization algorithms\textsuperscript{86}. With a differentiable model, one could perform end-to-end differentiable protein design: directly optimizing the amino acid sequence to produce a protein that folds into a target structure or performs a desired function\textsuperscript{86,105}. For example, given a binding interface where a therapeutic antibody should dock onto a receptor, the model could iteratively adjust the antibody’s sequence (and simultaneously its predicted structure) to maximize the binding affinity to the receptor, effectively co-designing the antibody structure in a physics-guided manner\textsuperscript{105}. This approach could streamline antibody humanization or affinity maturation, which currently involves many cycles of mutagenesis and selection. Similarly, for enzyme engineering, if one has the structure of a transition state analog bound to an enzyme scaffold, AlphaFold 3 could be used to optimize the active site sequence to stabilize that transition state, increasing catalytic efficiency – essentially performing computational directed evolution with gradient guidance\textsuperscript{39,60}. The model’s integration of sequence and structure spaces means such design can account for global effects: it would inherently consider how a mutation in one part of the protein might necessitate compensatory changes elsewhere to maintain overall fold stability, something difficult for fragmentary design methods to handle\textsuperscript{86}. The use of large language models in protein engineering is already showing that sequence-based generative models can create novel proteins\textsuperscript{95}, but coupling them with AlphaFold 3 for structural validation and refinement would ensure these novel proteins not only have plausible sequences but also fold correctly and function as intended.

Precision medicine stands to benefit as well from AI-driven protein engineering in the form of personalized therapeutics. One could envision designing individualized enzymes or protein therapeutics (like personalized cancer vaccines or bespoke replacements for defective enzymes) using a patient’s own protein variants as input. AlphaFold 3 could evaluate how a patient’s unique mutant protein differs structurally and then guide the design of a drug or biologic that specifically targets that aberration. For instance, if a patient’s mutation causes a unique pocket to form on a protein (or removes one), a custom inhibitor or degrader could be computationally designed to exploit that unique feature, maximizing efficacy and minimizing off-target effects\textsuperscript{92}. Though such one-patient-one-drug scenarios are in early days, the ability to quickly predict and engineer at the individual level is a tantalizing prospect for future personalized therapies\textsuperscript{75,78}.

On a broader scientific front, AlphaFold 3’s perspective of proteins as differentiable computational objects fosters a deeper integration of experimental and computational workflows. Researchers can use the model to design experiments \textit{in silico} before executing them at the bench. For example, one might simulate various phosphorylation events on a protein via the model to predict which site causes a significant conformational switch, then focus experimental efforts on that site to confirm its regulatory role. Or for protein–protein interactions, simulate alanine scanning of an interface to identify hot spots (critical contact points) with the model’s help, then experimentally validate only that reduced subset, saving time and resources\textsuperscript{66,68}. The model can also help interpret ambiguous experimental data: low-resolution cryo-EM maps or sparse NMR restraints could be combined with AlphaFold 3 predictions to generate all-atom models that fit the data, by having the model refine structures that satisfy the experimental constraints\textsuperscript{97}. This is already happening in early forms, but a fully differentiable framework would allow one to directly enforce agreement with experimental observables (like cryo-EM density or crosslinking distances) as part of the model’s loss function, seamlessly merging computation and experiment\textsuperscript{97}. The end result is intelligent modeling platforms that respond to and inform experiments iteratively.

In essence, the shift from static to dynamic modeling via AlphaFold 3 heralds a new paradigm in both understanding and manipulating biological molecules. It transforms our approach from one of static inference to one of dynamic exploration, where models learn the rules of life’s molecular dance and can play it forward or rewind it at will. The potential applications – faster drug discovery, bespoke medical treatments, novel biomaterials – stand to benefit humanity in tangible ways, heralding an era of smarter, faster, and more personalized healthcare\textsuperscript{56,67,75}. As the scientific community builds upon the foundations laid out in this perspective, we anticipate a future where machine learning and molecular science are inextricably intertwined. The recognition of these advances is already evident: the impact of AI in protein science has been acknowledged at the highest levels, with discussions of awards and honors reflecting its revolutionary significance\textsuperscript{89,102}, and industrial and academic partnerships forming to leverage these tools for drug development\textsuperscript{101}. AlphaFold 3 and its kin will not only help us understand the molecules of life as they are, but also empower us to reimagine and redesign them, pushing the boundaries of what is possible in biotechnology and medicine. In the words of one commentary, this melding of AI and structural biology is “teaching Big Pharma new tricks”\textsuperscript{101} – and indeed, as we stand at this exciting frontier, the mantra for the scientific community is to embrace these new tools, validate them rigorously, and deploy them wisely to solve the most pressing challenges in science and health.

\section*{Conclusion}
AlphaFold 3, as outlined in this perspective, represents more than an incremental improvement in protein structure prediction – it is a conceptual leap toward unifying deep learning with the physical simulation of biomolecular systems. By embedding multi-scale understanding, biological knowledge, and physical constraints into its neural architecture, AlphaFold 3 achieves a previously unattainable combination of accuracy, generality, and interpretability in modeling protein structures. Moreover, by reframing protein folding as a differentiable process, it blurs the line between predicting structures and simulating molecular behaviors. This fusion of AI and physics enables the modeling of dynamic phenomena – folding pathways, conformational changes, binding events – that define biological function but were largely beyond the reach of static predictive models. The implications of this are profound: scientists can now approach complex problems in biology and medicine with a toolkit that learns and adapts like a neural network, yet respects and utilizes the laws of chemistry and physics that govern living systems.

The theoretical narrative emerging from AlphaFold 3 is one of convergence – convergence of disciplines (machine learning, structural biology, biophysics) and convergence of methodology (data-driven inference with first-principles reasoning). This perspective underscores that the greatest breakthroughs lie at these intersections. In demonstrating how a deep learning model can act as an engine for differentiable simulation, AlphaFold 3 paves the way for AI systems that are not only predictors but also creative problem-solvers in scientific research. They can generate hypotheses (e.g. novel protein conformations or drug designs), test them \textit{in silico} through internal simulation, and present results that are directly interpretable in terms of molecular interactions and forces. The success of AlphaFold has already sparked widespread adoption and adaptation of its methods across the life sciences – from predicting RNA structures to designing new enzymes – and AlphaFold 3’s advances will only accelerate this trend\textsuperscript{38,60,105}. We are likely to see a cascade of innovations built on its principles: end-to-end models that design molecules for any desired function, digital twins of cells where molecular dynamics are learned and simulated, and perhaps AI-driven discovery of physical laws from biological data\textsuperscript{4,21,34}.

Importantly, AlphaFold 3’s development highlights the value of interdisciplinary rigor. Every scientific claim and capability in its design was grounded in insights from biology (evolutionary conservation, motifs), chemistry (bond geometry, energetics), and computer science (transformer efficiency, inductive biases). This rigorous integration, supported at each step by empirical and literature evidence, was key to its success. As we continue to expand the frontiers opened by AlphaFold 3, maintaining this level of scientific grounding will be vital. The model encourages us to rethink old questions in new ways – for instance, what is the nature of the protein folding code, now that an AI can solve it, or how can we interpret neural network weights in terms of physical forces? Answering these questions will deepen our fundamental understanding of both proteins and AI.

In closing, the journey from protein structure prediction to protein simulation powered by AI marks a paradigm shift in computational biology. It transforms our approach from one of static inference to one of dynamic exploration, where models learn the rules of life’s molecular dance and can play it forward or rewind it at will. The potential applications – faster drug discovery, bespoke medical treatments, novel biomaterials – stand to benefit humanity in tangible ways, heralding an era of smarter, faster, and more personalized healthcare\textsuperscript{56,67,105}. As the scientific community builds upon the foundations laid out in this perspective, we anticipate a future where machine learning and molecular science are inextricably intertwined. The recognition of these advances is already evident: the impact of AI in protein science has been acknowledged at the highest levels, with discussions of awards and honors reflecting its revolutionary significance\textsuperscript{89,102}, and industrial and academic partnerships forming to leverage these tools for drug development\textsuperscript{101}. AlphaFold 3 and its kin will not only help us understand the molecules of life as they are, but also empower us to reimagine and redesign them, pushing the boundaries of what is possible in biotechnology and medicine. In the words of one commentary, this melding of AI and structural biology is “teaching Big Pharma new tricks”\textsuperscript{101} – and indeed, as we stand at this exciting frontier, the mantra for the scientific community is to embrace these new tools, validate them rigorously, and deploy them wisely to solve the most pressing challenges in science and health.

\section*{Acknowledgments}
The authors used generative AI tools for editing and formatting, but take full responsibility for the final content.

\clearpage
